\documentclass[sigconf]{acmart} \usepackage{graphicx}
\usepackage{subcaption}
\usepackage{multirow}
\usepackage[linesnumbered,ruled,vlined]{algorithm2e}
\usepackage{float}
\usepackage{color}
\usepackage{booktabs}
\usepackage{enumitem}
\usepackage{balance}
\usepackage{arydshln}

\definecolor{pro_green}{rgb}{0.0, 0.66, 0.47}

\newcommand{\eat}[1]{}

\newcommand{\ipcs}{index-based point cloud color swapping\xspace}
\newcommand{\rr}{restricted rendering-based\xspace}

\newcommand{\defense}{$IPC^{2}S$ defense\xspace}
\newcommand{\renderingdefense}{$R^2$ defense\xspace}
\newcommand{\rec}{recognition\xspace}

\newcommand{\para }[1]{\noindent  {\bf #1}}
\newcommand{\1}{{\em (i)}}
\newcommand{\2}{{\em (ii)}}
\newcommand{\3}{{\em (iii)}}

\usepackage{breakurl}
\PassOptionsToPackage{hyphens}{url}\usepackage{hyperref}

\hypersetup{
    colorlinks=true,
    linkcolor=magenta,
    filecolor=magenta,      
    urlcolor=magenta,
    citecolor=magenta,
}

\copyrightyear{This is the authors' version of the work. 2022}
\acmYear{2022}
\setcopyright{acmcopyright}
\acmConference[MM '22] {30th ACM International Conference on Multimedia}{October 10--14, 2022}{Lisbon, Portugal.}
\acmBooktitle{30th ACM International Conference on Multimedia (MM '22), October 10--14, 2022, Lisbon, Portugal.}
\acmPrice{15.00}
\acmDOI{10.1145/XXXXXX.XXXXXX}
\acmISBN{978-1-4503-7988-5/20/10}

\acmSubmissionID{2933}

\begin{document}
\title{Privacy-preserving Reflection Rendering for Augmented Reality}

\author{Yiqin Zhao}
\affiliation{\institution{Worcester Polytechnic Institute}
  \city{Worcester}
  \state{MA}
  \country{USA}}
\email{yzhao11@wpi.edu}

\author{Sheng Wei}
\affiliation{\institution{Rutgers University}
  \streetaddress{94 Brett Road}
  \city{Piscataway}
  \state{NJ}
  \country{USA}}
\email{sheng.wei@rutgers.edu}

\author{Tian Guo}
\affiliation{\institution{Worcester Polytechnic Institute}
  \city{Worcester}
  \state{MA}
  \country{USA}}
\email{tian@wpi.edu}

\begin{abstract}
Many augmented reality (AR) applications rely on omnidirectional environment lighting to render photorealistic virtual objects. 
When the virtual objects consist of reflective materials, such as a metallic sphere, the required lighting information to render such objects can consist of privacy-sensitive information that is outside the current camera view. 
In this paper, we show, for the first time, that accuracy-driven multi-view environment lighting can reveal out-of-camera scene information and compromise privacy. 
We present a simple yet effective privacy attack that extracts sensitive scene information such as human face and text information from the rendered objects, under a number of application scenarios.

To defend against such attacks, we develop a novel \defense and a conditional \renderingdefense. 
Our \defense, used in conjunction with a generic lighting reconstruction method, preserves the scene geometry while obfuscating the privacy-sensitive information. 
As a proof-of-concept, we leverage existing OCR and face detection models to identify text and human faces from past camera observations and blur the color pixels associated with detected regions.
We evaluate the visual quality impact of our defense by comparing rendered virtual objects to ones rendered with a generic multi-lighting reconstruction technique, ARKit, and \renderingdefense. 
Our visual and quantitative results demonstrate that our defense leads to structurally similar reflections with up to 0.98 SSIM score across a variety of rendering scenarios while preserving sensitive information by reducing the automatic extraction success rate to at most 8.8\%. 

\end{abstract}

\ccsdesc[500]{Security and privacy~Systems security}
\ccsdesc[500]{Information systems~Multimedia streaming}

\keywords{Augmented reality; visual privacy; photorealistic rendering}

\begin{teaserfigure}
\centering
    \includegraphics[width=0.9\textwidth]{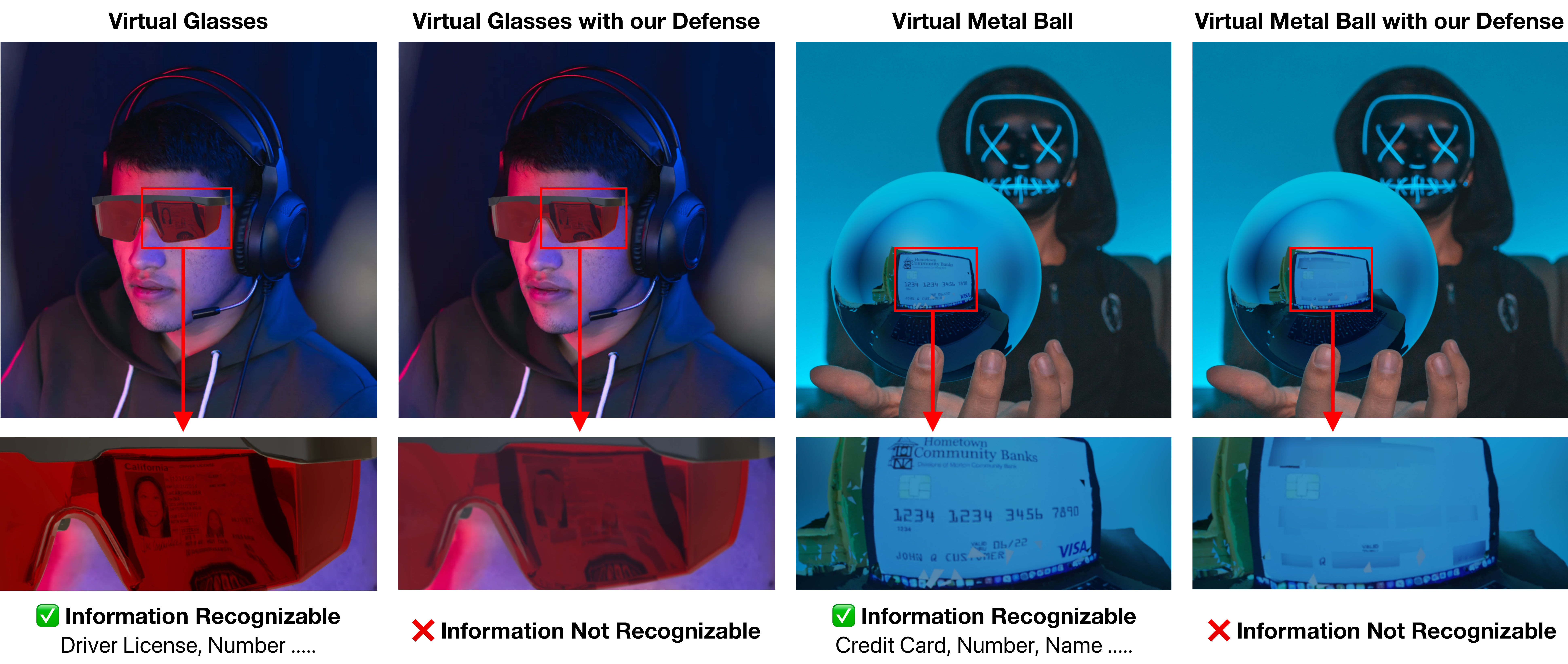}
    \caption{
    Sensitive information on reflection rendering with/without our defense.
        \textnormal{
        In the 1st and 3rd columns, sensitive information from the physical environment can appear as part of the rendered reflections and be leaked to viewers. 
        In the 2nd and 4th columns, we show that our proposed defense can effectively eliminate such information leakage while still keeping high visually coherent reflections.
}
    }
    \label{fig:teaser}
\end{teaserfigure}

\maketitle

\section{Introduction}

Augmented reality (AR) has the promise to transform many aspects of our lives, including education~\cite{Google_for_Education_undated-rj}, healthcare~\cite{Andrews2019-qh}, and business~\cite{Inter_IKEA_Systems_B_V2017-iv}. By 2023, mobile AR is predicted to be a hundred-billion dollar market, with hundreds of millions of users~\cite{Alsop2020-lg}. 
Today, a new form of AR application is increasingly supported by many popular social media apps such as TikTok and YouTube.
This new application scenario, which we refer to as \emph{AR content creation/streaming}, allows users to create videos augmented with 2D/3D assets~\cite{vroid_studio}.
The AR content can then be shared via social media platforms. 
For example, a streamer may be interested in engaging her community with a virtual sunglasses try-on session where she will try on different sunglasses based on the text chat suggestions from the community.

This engagement-driven AR content creation often calls for good visual effects, which further translates to the high rendering quality of the virtual objects. 
That is, the virtual objects appearing in the video stream should exhibit \emph{visual coherency} to the physical world background and should be rendered in a photorealistic way.
To achieve good visual effects, it is desirable for AR applications to have omnidirectional environment lighting. 
To provide accurate lighting information, existing AR frameworks often require AR users to scan the physical environment and leverage deep learning models to estimate the lighting~\cite{arcore,arkit,zhao2022FusedAR}. 
Without loss of generality, we refer to this type of lighting estimation approach as \emph{multi-view lighting reconstruction}.

However, environment scanning will capture multiple glimpses (i.e., camera views) of the physical environment, some of which can consist of privacy-sensitive information and be \emph{out-of-camera} scene during the streaming session.
The privacy problem is further exacerbated with the advent of 3D vision sensors (e.g., LiDAR). 
This newly endowed capability to mobile devices is a double-edged sword: it allows mobile devices to more efficiently capture and accurately reconstruct physical scenes for better AR features; it also presents an immediate threat in a new form of \emph{reflection-based privacy}. 
Figure~\ref{fig:teaser} demonstrates two examples where sensitive information such as a driver's license and a credit card can appear in reflective virtual objects.
Consequently, when streaming AR content consisting of such objects (as will show in Figure~\ref{fig:AR_streaming_workflow}), it can lead to undesired information leakage to any streaming viewers, without the AR streamers necessarily noticing. 

In this paper, we show, for the first time, that visual quality-driven multi-view lighting reconstruction can reveal out-of-camera scene information and compromise privacy for AR content creators.
Existing works that support reflective rendering, including commercial methods in ARKit~\cite{arkit} and academia research GLEAM and FusedAR~\cite{prakash2019gleam,zhao2022FusedAR}, all require the step to capture multiple glimpses of the physical environment. 
Without loss of generality, we present a privacy attack, based on a recently proposed lighting reconstruction technique FusedAR~\cite{zhao2022FusedAR}, that extracts sensitive scene information such as human face and text information from the rendered objects, under a number of plausible application scenarios.
One of our key goals in demonstrating the effectiveness of this simple attack is to \emph{increase the awareness of privacy issues associated with reflection rendering for AR applications}. 

We note that visual privacy protection is not a new problem~\cite{Cho-ICPR08,Patel-ppvs09}. Prior work has proposed a plethora of defenses for traditional multimedia, such as images and videos~\cite{Zhu-mobicom17,Wu-mobicom21}. 
Even for emerging mixed reality applications, we have also observed increased research efforts to ensure that an immersive virtual environment is built with security and privacy implications in mind~\cite{Shi-mobicom21,Luo-NDSS20,Luo-VR22}.
Our paper falls into the broad AR/VR privacy research; one of our main contributions is \emph{uncovering this new reflection-based privacy issue in the emerging AR applications}.  
We argue that the demonstrated attack is a natural progression from the improvement of mobile sensors and environment understanding algorithms~\cite{Gardner2017,Song2019,hdr-environment-map-estimation}. 
In other words, this reflection rendering-based attack is a consequence of improved lighting reconstruction for AR applications.

To defend against such attacks, we develop a novel privacy-preserving \defense that \emph{preserves the geometry information} while obfuscating the privacy-sensitive objects. 
Preserving the geometric information is critical in addressing the key challenge of simultaneously preserving privacy while still delivering visually coherent rendering.
Additionally, we propose a \renderingdefense that can bypass the lighting reconstruction and provide effective protection in dynamic environment conditions such as low lighting or motion blur. 
We leverage existing OCR and face detection models~\cite{easyocr,opencv_library} to identify text and human faces from past camera observations and blur the \emph{color pixels} associated with detected regions.
The transformed RGB images with the \emph{unmodified depth information} are further combined into a point cloud, a 3D intermediate data we use to generate the final environment map for rendering.

To demonstrate that our \defense can successfully obfuscate private information while delivering good visual effects, we evaluate the defense pipeline under 32 different rendering scenarios.
We show that our \defense achieves high visual quality with up to 36db PSNR and 0.98 SSIM while significantly reducing the automatic extraction success rate from 97.1\% to 8.8\% when compared to the privacy-risking reflection renderings.
Lastly, we find that in addition to the three factors---physical scene, virtual object, and sensitive information, the accuracy of the face and text \rec models also can impact the information extraction success rate and the visual quality. 
We make the following main contributions.
\begin{itemize}[leftmargin=.12in,topsep=0pt]
    \item We present the first look at the \emph{out-of-camera visual privacy} issue, i.e., the \emph{reflection-based privacy}, that arises in AR applications. 
    To demonstrate the prevalence of the privacy issue, we showcase a multi-view attack based on the ARKit and a recent lighting reconstruction technique~\cite{zhao2022FusedAR} that successfully extracts sensitive information from reflective virtual objects.
    \item We propose an effective \defense to automatically remove sensitive information from user-defined categories, such as human faces or textual information, and a conditional \renderingdefense. Our \defense is a lightweight pipeline that leverages machine learning models and image blurring techniques and can run in parallel to the current reflective rendering in AR frameworks.
    \item We implement the pipelines for lighting reconstruction, attack, and defense, and conduct evaluations to compare both visual quality and sensitive information extraction. Relevant research artifacts are available at \textcolor{magenta}{\url{https://github.com/cake-lab/ar-reflection-privacy}}.
\end{itemize}

\section{Background}

\begin{figure*}[t]
    \centering
    \includegraphics[width=0.9\textwidth]{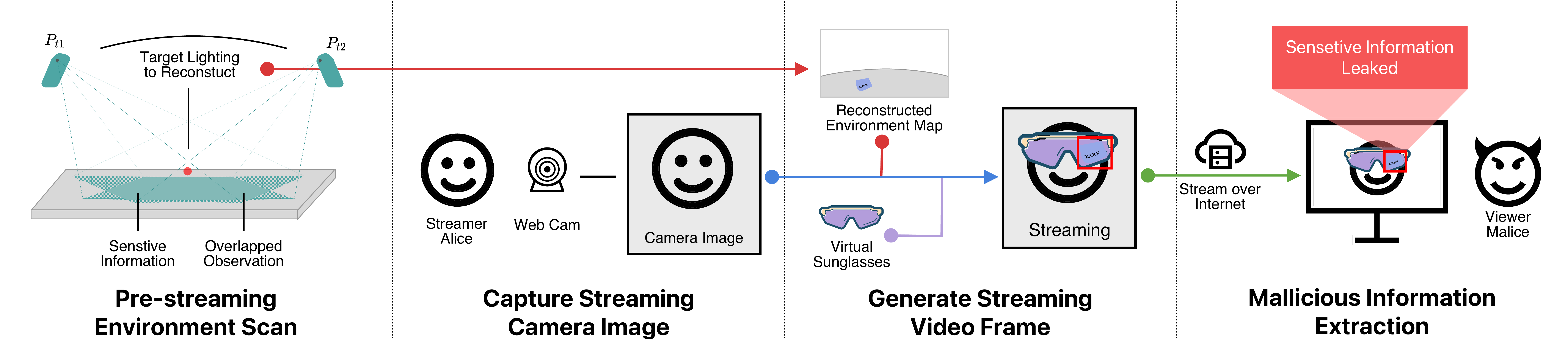}
    \caption{The reflection-based privacy issues in AR content creation and streaming.
    \textnormal{We demonstrate the general workflow of AR streaming to highlight how the rendered reflection of virtual sunglasses can contain sensitive information from the streamer Alice's physical surroundings and be leaked to the viewer Malice.
    }
    }
    \label{fig:AR_streaming_workflow}
    \vspace{-4mm}
\end{figure*}
 
\para{Virtual Object Rendering in AR Streaming.}
Imagine a content streamer Alice that uses AR-enabled applications, such as TikTok, to stream using platforms such as Twitch~\cite{twitch}.
In this case, Alice is interested in augmenting her video stream with rendered objects. 
Depending on the streaming scenarios, e.g., virtual try-on of glasses or furniture shopping, Alice also would like to have visual-coherent virtual objects overlay in her physical space.  
To produce visual-coherent virtual objects, the AR frameworks will need to have access to accurate environment lighting information~\cite{prakash2019gleam,pointar_eccv2020}.
The current commercial AR frameworks such as ARKit or ARCore often require mobile users to move around the cameras to scan the physical environment~\cite{arkit,arcore}. 
The scanning phase will allow AR frameworks to collect useful environment information, which will further be used as input for a lighting estimation module to output lighting information~\cite{Somanath2020-of}.
The environment lighting information, often represented in the form of \emph{environment map}~\cite{debevec2006image}, will then be used by rendering frameworks to overlay the virtual objects either in a user-specified world position~\cite{pointar_eccv2020} or a position based on tracking results~\cite{Tang2021-ul}.
Finally, each video frame augmented with virtual objects can be piped to existing streaming software such as OBSStudio~\cite{obs} to use services like Twitch.

\para{Visual Privacy Considerations in AR.}
Considering the AR streaming scenario described above, we will next describe scenarios that will lead to privacy issues. 
The key privacy problem arises when the AR frameworks use captured environment images as part of the input for reconstructing environment lighting information~\cite{prakash2019gleam,zhao2022FusedAR}.
These environment images captured during the environment scanning phase (before the streaming) can lead to \emph{out-of-camera} information leak when the AR framework uses a high-quality environment map to render reflective objects.
We refer to this privacy problem as \emph{reflection-based privacy} in which sensitive information from outside the current camera frame can appear on the virtual objects.  
Figure~\ref{fig:vis_attack} shows example rendering effects of our proposed simple attack that leverages a popular AR framework ARKit and a recently proposed multi-view lighting reconstruction method~\cite{zhao2022FusedAR}.
When streaming the augmented video frames, such sensitive information will then become accessible to any viewers over the internet. 
More generally, such privacy issues can happen in many AR applications that satisfy the following characteristics.
\1 \emph{The need for photorealistic rendering.} Many compelling use cases of AR require photorealistic rendering. For example, in a 3D advertisement where an influencer tries to sell products (as rendered assets) to followers.
\2 \emph{Physically separated users.} While many AR applications are multi-user, we have observed scenarios where AR users record and share their experiences via various platforms like Snapchat. In such scenarios, the existing platform users do not have to engage in AR technology directly but rather as consumers of AR content (see Figure~\ref{fig:AR_streaming_workflow}).

\section{Lighting Reconstruction Premier} 
\label{sec:lighting_reconstruction_premier}
In this section, we present the critical information of \emph{multi-view lighting reconstruction} (refer to Figure~\ref{fig:reflection_extraction}) which serves as the basis for the privacy issues we pinpoint in \S~\ref{sec:attack} and defenses in \S~\ref{sec:defense}.

\para{Step 1: Capturing Environment Data.}
Most mobile devices only have cameras with relatively small field-of-view, e.g. 77$^\circ$~\cite{pixel5specs}. Therefore, to capture omnidirectional environment observations, AR content creators are typically required to move the mobile device around and scan the surroundings.
Traditionally, the capturing can be performed with the assistance of a physical chrome ball~\cite{prakash2019gleam,debevec2006image}. In recent years, the increasingly popular mobile depth sensor~\cite{Huawei_undated-ca,ipad-lidar} enables the possibility of capturing highly accurate scene geometry. 
Similar to recent work~\cite{zhao2022FusedAR}, we perform lighting reconstruction with RGB-D images and device tracking data captured by mobile devices without requiring additional scene setups.

\para{Step 2: Combing Multi-View Data.}
Next, we combine the captured multi-view data into a 3D point cloud representation in the same world space.
We select the point cloud based on the virtual object rendering position within a cubic space with a size of 2 meters as \emph{near-field}.
To ensure the reconstruction quality, we only select the points with high depth confidence values, which measures the accuracy of the depth-map data.
Moreover, we perform view-wise point cloud registrations using iterative closest point registration~\cite{besl1992method} to address noisy real-world tracking data.

\para{Step 3: Finalizing Environment Lighting.}
Last but not least, we convert the collected near-field point cloud into an environment map, which is composed of near and far-field components: 
\1 The near-field component consists of the projection of the textured surface mesh reconstructed from the collected point cloud;
\2 for the far-field component, we use an indoor blurred HDR panorama image, similar to the far-field reconstruction policy described in~\cite{zhao2022FusedAR}.

\section{Reflection-based Privacy Issues}
\label{sec:attack}

\subsection{Privacy Attack Overview}
\label{subsec:attack}

Figure~\ref{fig:AR_streaming_workflow} presents an overview of the AR streaming workflow where reflective rendering can lead to creators' physical environment information being recovered by viewers.  
AR streaming, as we defined previously, is an emerging and popular form for indie streamers to reach out to followers via platforms such as Tiktok, YouTube, and Twitch~\cite{Lu2021-wi,Tang2021-ul}.
We assume that streamers are using existing AR software to create videos with seamlessly overlayed virtual objects.

To generate quality content, e.g., photorealistic rendered objects that are coherently inserted into the physical scenes, our streamer Alice often needs to use the camera device to scan her physical surroundings. 
This step of \emph{environment scanning} is a basic requirement of existing commercial AR frameworks such as ARCore or ARKit to obtain useful AR information including world tracking data, camera intrinsic, and camera pose. 
As a consequence of this scanning, the AR session (and subsequently the AR stream) will have access to the physical world information surrounding Alice. 
Note that, we assume that such physical world information, if not \emph{directly captured} by the camera during the streaming, should not be available to viewers. 
However, in the AR streaming case, when rendering virtual objects with reflective materials, e.g., the streamer wants to show the viewers the sunglasses try-on experience, the virtual sunglasses will be rendered with environment lighting information that was captured previously.
Simply put, the virtual sunglasses might reflect different sensitive information, such as human faces or credit card information, when the streamer looks around (recall Figure~\ref{fig:teaser}). 
Finally, the attacker, i.e., the AR content viewer, can access the rendered images or video clips shared by the AR content streamers directly on the attacker's device.

\subsection{Sensitive Information Extraction}
\label{subsec:high_frequency_lighting}

\begin{figure*}[t]
    \centering
    \includegraphics[width=0.9\textwidth]{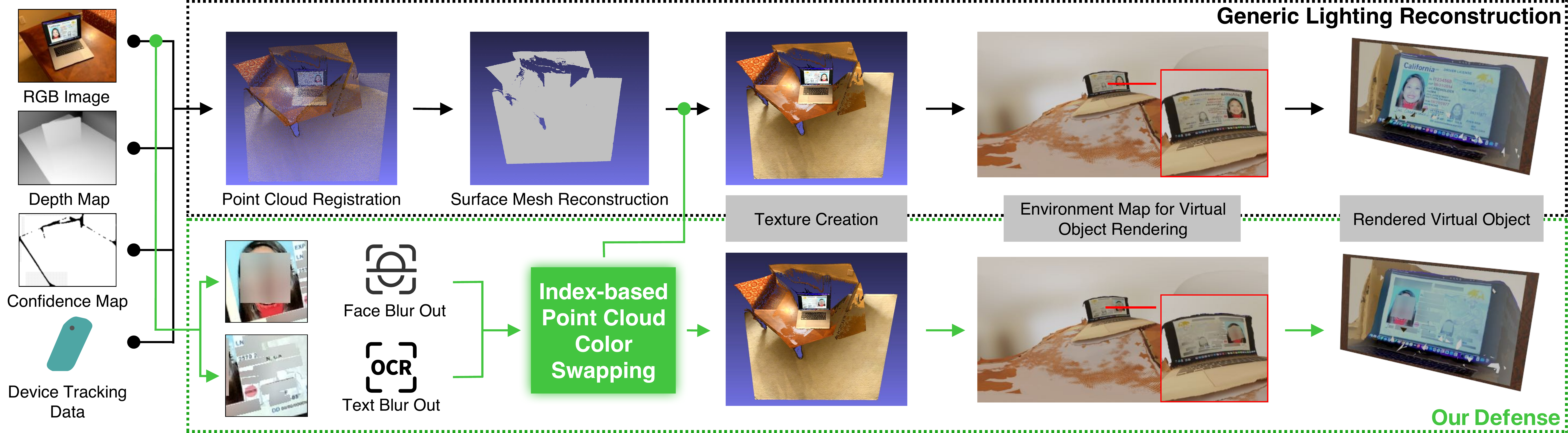}
    \caption{
    Our proposed defense for preserving reflection-based privacy for AR streaming. 
    \textnormal{Our defense pipeline is designed around the key idea of index-based point color swapping to obfuscate the visual detail of the sensitive information while still maintaining the overall reflection color pattern.
    Moreover, the defense pipeline can run in parallel to the generic lighting reconstruction pipeline described in Sec~\ref{sec:lighting_reconstruction_premier}.
    }
    }
    \label{fig:reflection_extraction}
    \vspace{-3mm}
\end{figure*}
 
To demonstrate the prevalence of privacy issues in reflection rendering, we design a simple attack and show that we can successfully extract sensitive information from the rendered reflections.
Our attack first reveals the out-of-camera sensitive information captured during lighting reconstruction by obtaining reflective virtual objects. 
For example, the attacker could ask the streamer to use existing reflective objects or hack the model assets used by the streamers and insert reflective components.
Then, we perform automated sensitive information extraction by running face and OCR \rec models.
Optionally, the attacker can unwrap the reflection area based on either virtual object geometry, viewing perspective, or both, if known.
The unwrapping step will further increase the chance of automated information extraction as it removes the projection distortion.

In \S~\ref{subsec:attack_effectiveness}, we demonstrate that the automated attack can achieve a face/text extraction success rate of 63.24\%/57.44\%, on average, across a diverse set of rendering scenarios.
In summary, given the ease of attack and the effectiveness in extracting sensitive information, we argue the rising need to protect AR content creation applications. 
Content creators might not realize that the visual information of their physical environment is being utilized for virtual object rendering. 
Such information leakage is unintended and undesirable and we believe it is necessary to have an automatic pipeline to identify when privacy concerns arise and to provide robust mechanisms to minimize unintended privacy leakage.

\section{Privacy-preserving Reflection}
\label{sec:defense}

We present the design of two defense mechanisms that effectively and efficiently protect \emph{reflection-based privacy}.
Specifically, we design \emph{\ipcs}, an automatic defense method we refer to as \defense, for face and text information.
As shown in Figure~\ref{fig:reflection_extraction}, \defense is designed to run in parallel to the lighting reconstruction supporting reflection rendering and addresses the visual privacy issue by blurring out the sensitive information fields.
We further extend our defense design to support dynamic environments and propose a \rr method (referred to as \renderingdefense) to still provide privacy protection when the automatic defense \renderingdefense falls short.

\subsection{Key Design Challenge and Questions}

Recall that the \emph{reflection-based privacy} issue arises as AR frameworks strive to improve the visual coherency of the AR content.
The key challenge when designing the defense is to minimize the impact on user-perceived visual quality while providing robust privacy protections for AR content creators.
We tackle this challenge by answering two key design questions. 
\emph{First, how to integrate the defense with the existing AR pipelines (\S \ref{subsec:ipccs})}? 
To generate visually coherent environment lighting for rendering virtual objects, existing AR frameworks such as ARKit often need to transform the environment scene observations multiple times, based on viewing perspectives and scene geometries.
It is therefore critical to carefully choose \emph{where and how to} perform the image obfuscation for visual privacy without introducing visual artifacts or performance overhead.
\emph{Second, how to ensure privacy protection in dynamic environments (\S~\ref{sec:rendering_defense})?}
It is common for AR content creators to work in dynamic environments, e.g., with a rapid change of environment lighting or moving cameras.
These environments can be challenging for automatic defenses using deep learning models~\cite{Liu2020-fy}.

\subsection{Index-based Point Cloud Color Swapping}
\label{subsec:ipccs}

To address the reflection-based privacy issue via image obfuscation, there are three main locations in the lighting reconstruction pipeline one can choose to obfuscate:
\1 reconstruction client device camera RGB images; 
\2 reconstructed lighting environment maps; and
\3 the rendered virtual object frames.
However, neither \2 nor \3 are ideal as they are subject to image distortion from the panoramic image projection and geometry/viewing-perspective-related distortion in rendering, respectively.
Instead, our \defense pipeline takes the input directly from the RGB images collected from mobile devices as such images do not suffer from lighting reconstruction and rendering-related distortions.
By doing so, we can avoid the negative impact of image distortion on \rec~\cite{Liu2020-fy}.

However, such a design comes with its own unique challenge. 
That is, obfuscating the RGB images at the early stage of lighting reconstruction can impact the lighting reconstruction accuracy, as the obfuscated pixel color information could be used in point cloud registration.
To address this challenge, we propose the \defense, a novel design that allows identifying privacy issues at the early lighting reconstruction stage while \emph{waiting to obfuscate} the privacy content in a later stage.
The key idea is to correctly and efficiently map the pixels of sensitive information to the points in the intermediate point cloud.

As shown in Figure~\ref{fig:reflection_extraction}, parallel to the main reconstruction process, we spawn a separate process that first generates blurred RGB images for each RGB image received by the lighting reconstruction pipeline.
Then, in the defense execution process, we run face and text detection models to \rec the information fields that appeared on each RGB image and use a bounding box to describe the information field regions.
Next, the identified regions are recorded as pixel and frame number indexes, which correspond to the index of points in the intermediate point cloud of the lighting reconstruction. 
At the later stage of lighting reconstruction (i.e., prior to mesh texturing), we swap the previously identified sensitive information regions based on point cloud indexes.
This design enables both high accuracy face/text \rec and eliminates the impact of image obfuscation on lighting reconstruction quality.

Furthermore, the \defense runs in parallel to the unmodified lighting reconstruction pipeline, which has the potential to minimize the latency impact of \defense and to be integrated with other reconstruction pipelines.
Empirically, our measurements show an average execution time of face/text \rec at 0.05s/0.13s compared to the 30s needed by the point cloud registration.
More concretely, this means that the high-quality lighting information (which is used to render reflection) can take up to 30 seconds using the reference pipeline implemented based on FusedAR~\cite{zhao2022FusedAR}. Our defense pipeline does not pose a performance bottleneck on normal AR usage as its main steps are complete well before the lighting information is ready.

\vspace{-3mm}
\subsection{Defense in Dynamic Environments}
\label{sec:rendering_defense}

Besides image distortion, the automatic face/text \rec accuracy can be affected by other environmental factors like lighting, camera movements, etc.
For example, in AR content creation and streaming applications, lighting reconstruction is usually performed in two cases: \emph{prior to the beginning of streaming} and \emph{during the streaming}.
Automatic face/text detection failures caused by sudden environmental changes in the pre-streaming scenario may not cause immediate privacy issues as AR content creators can be given an opportunity to re-scan the environment to avoid sharing sensitive information with the viewers.
However, it is important that our defense handles the dynamic environment during the AR streaming as \rec failures will lead to imminent privacy leakage. 

Therefore, in addition to \defense, we propose the \renderingdefense which bypasses the lighting reconstruction and can provide immediate protection.
\renderingdefense controls the maximum material reflection rate and roughness, which can be executed easily and efficiently in most modern graphics rendering engines.
In particular, we limit the maximum material reflection to 0.8 (from 1.0) and minimum roughness to 0.2 (from 0.0). 
The changing environments can be detected by leveraging built-in hardware sensors, e.g. ambient light sensor, accelerometer, and gyroscope.

\section{Experiments}

We evaluate our proposed defense's effectiveness in preserving privacy and the respective impact on the rendering quality of virtual objects.
Our evaluation centers around answering the key question: \emph{how well does our defense work in preserving privacy and maintaining good visual coherency?}
We test a total of 32 rendering scenarios, where each scenario refers to an instance of \emph{(scene, reflective object, sensitive information)}, for both information extraction and rendering quality. 
Our key findings are summarized as follows: 
\begin{itemize}[leftmargin=.12in,topsep=0pt]
\item Our simple attack can successfully extract up to 100\% human faces and at least 92\% textual information when inspecting the extractions manually (\S~\ref{subsec:attack_effectiveness}). 
\item Our \defense effectively reduces the information extraction success rates to at most 8.8\% and 23.8\% under automatic and manual inspections, for all tested rendering scenarios. (\S~\ref{subsub:defense_success_rate}).  
\item Compared to the \renderingdefense, our \defense achieves an average of 9.31\% better SSIM while successfully preserving privacy under manual inspection(\S~\ref{subsub:defense_visual}).
\item Automatic extraction poses more difficulty for identifying faces than texts, for both the attack and the defense.
\end{itemize}

\vspace{-4mm}
\subsection{Experimental Setup}
\label{sec:exp_setup}

\para{Implementations.} To demonstrate the prevalence of the reflection-based privacy issues in AR, we implement \emph{a simple lighting reconstruction pipeline}, following the generic lighting reconstruction pipeline paradigm, by leveraging widely available open-source tools and libraries.
The lighting reconstruction pipeline consists of both an iOS app developed with Unity3D~\cite{unity3d} and the ARFoudnation framework~\cite{arfoundation} as the client and a Python server. 
During reconstruction, we first stream the collected AR scene information, RGB-D image, device tracking data, and camera pose, from the mobile client to the backend server and store the scene information for further processing.
Next, we perform the point cloud registration and surface reconstruction on the backend server using the Open3D~\cite{Zhou2018} and the Meshlab~\cite{cignoni2008meshlab} libraries.
Finally, we generate environment maps from reconstructed meshes using Blender~\cite{blender3d}.

We use the reconstructed environment map to implement \emph{our attack} by first rendering reflective objects using Blender.
Then, we implement the automatic sensitive information extraction of face and text \rec with OpenCV~\cite{opencv_library} and EasyOCR~\cite{easyocr} libraries.
We envision the attack will occur as a natural progression of AR frameworks supporting reflective rendering---\emph{malicious users/viewers do not need to investigate the inner-working of the pipeline}; rather, malicious users simply need to gain access to the rendered reflections.
We choose to implement our defense pipeline as a parallel component to the generic lighting reconstruction pipeline as described in the FusedAR paper~\cite{zhao2022FusedAR}.
Using the collected RGB image during lighting reconstruction, we use the same face and text \rec tools for the defense (as the attack for a fair comparison) to identify sensitive information and blur sensitive information of the rendered images using the PIL framework~\cite{umesh2012image}.
The \defense is implemented as a NumPy~\cite{harris2020array} ndarray operation and the  \renderingdefense as a special material in Blender.

\para{Rendering Scenarios.}
We first use an iPad Pro with a LiDAR sensor to capture RGB-D images and device tracking information in \emph{four indoor scenes} with different scene geometries and physical objects and reconstruct the environment lighting for each scene.
We choose two virtual objects with representative geometries, a \emph{metallic sphere} and a \emph{flat mirror}, to render the reflection.  
For the sensitive information, we select two sample US driver's licenses (Massachusetts and California), one \emph{group photo} with 14 persons, and one sample \emph{credit card} with 7 information fields, to represent three types of information leakage---a mixed of text and human face, human face-only, and text-only, respectively. 
\emph{Driver's license 1} contains a total of 19 (1 face and 18 texts) information fields and the \emph{driver's license 2} contains a total of 19 (2 faces and 17 texts) information fields.
For simplicity, we display the sensitive information on a screen of a Macbook Pro 15" inside each scene. 

At a high level, the environment capturing process involves a user scanning the indoor scene that consists of sensitive information. 
The depth map is captured at the resolution of 256x192, and the RGB color image is captured at the resolution of 1280x960.
We then import the reconstructed environment mesh into Blender to generate an environment cubemap at the resolution of 2048x2048 per cube face.
The environment map is composed of near and far-field components: 
\1 The near-field component consists of the projection of the near-field textured scene geometry;
\2 for the far-field component, we use an indoor blurred HDR panorama image, similar to the far-field reconstruction policy described in~\cite{zhao2022FusedAR}.
Note that all of our attack and defense evaluations are based on the near-field geometries---any panorama image can be used for the far-field without impacting the observed results. 
With the generated environment maps, we use Blender with the Principled BSDF shader~\cite{blender_bsdf} to render the virtual objects which will then be displayed.

\para{Evaluation Baselines and Metrics.}
We choose a commercial AR framework ARKit~\cite{arkit} and a recent lighting reconstruction pipeline FusedAR~\cite{zhao2022FusedAR} as the baselines of visual quality. 
To evaluate whether the proposed attack can successfully extract the sensitive information, i.e. the effectiveness of the attack, we use the success rate of information extraction as our metric. 

\vspace{-2mm}
$$
\textit{success rate} = \frac{\# \text{ of successful cases of information extraction}}{\# \text{ of valid test cases}}
$$

For each detected text field, we calculate the \emph{Levenshtein distance} between the recognized value and its ground truth value. 
A \rec is considered successful if the following conditions are both met:
\1 the recognized value is not empty; and
\2 the Levenshtein distance is less than 10.
For face \rec, we draw face bounding boxes on the reflection renderings and manually inspect whether each face is detected or not.

Furthermore, we use Peak signal-to-noise ratio (PSNR) and Structural Similarity Index (SSIM), commonly used image quality metrics, to quantify the impact of our defense on perceptible visual quality.
Both PSNR and SSIM values are calculated by comparing the virtual objects rendered with and without the proposed defense.

\subsection{Information Extraction of Our Attack}
\label{subsec:attack_effectiveness}

\begin{figure*}[t]
    \centering
    \includegraphics[width=\textwidth]{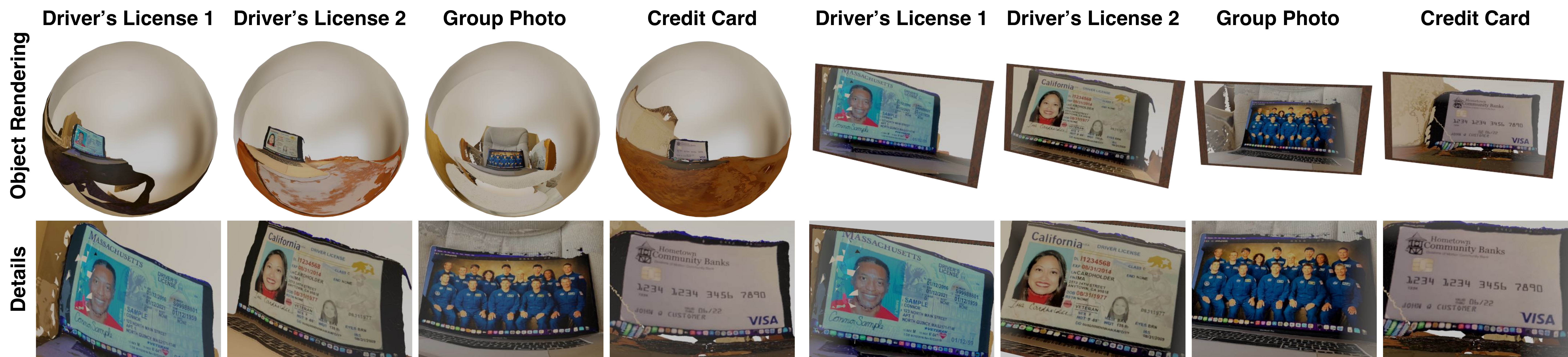} \vspace{-5mm}
    \caption{
        The visual effectiveness of our attack on reflection-based privacy.
        \textnormal{
        Row one shows the reflection rendering of two virtual objects;
        row two zooms in on the four types of sensitive information leakage.
        }
    }
    \label{fig:vis_attack}
    \vspace{-3mm}
\end{figure*}

We evaluate the effectiveness of our proposed attack by attempting to extract sensitive information from \emph{images of reflective virtual objects} rendered with reconstructed environment lighting.
Using environment lighting reconstructed from 4 scenes, we generate 8 renderings of both metallic ball and mirror objects for each scene.
We evaluate the success rates of both automatic extraction and manual inspection methods of a total of 17 faces and 42 text fields that appear in the reflection renderings.

\setlength{\tabcolsep}{3pt}
\begin{table}[t]
\centering
\caption{The high success rate of our attack.
\textnormal{On average we can manually extract 100\% human face information and at least 92\% textual information.
The automatic extraction has lower success rates but still poses considerable privacy issues.
}
}
\vspace{-2mm}
\small{ \begin{tabular}{lrrrrr}
\toprule
\textbf{Sensitive} & \textbf{Virtual} & \multicolumn{2}{r}{\textbf{Face Recognition}} & \multicolumn{2}{r}{\textbf{Text Recognition}} \\ 
\textbf{Info} & \textbf{Object} & \textbf{\textit{Automatic}} & \textbf{\textit{Manual}} & \textbf{\textit{Automatic}} & \textbf{\textit{Manual}}\\

\midrule

Driver's                & Metal Ball    & 50.00\%  & 100.00\% & 33.33\% & 87.50\% \\
License 1   & Mirror        & 100.00\% & 100.00\% & 81.94\% & 97.83\% \\

\midrule

Driver's                & Metal Ball    & 100.00\% & 100.00\% & 20.59\% & 94.12\% \\
License 2   & Mirror        & 100.00\% & 100.00\% & 80.88\% & 95.59\% \\

\midrule

Group                   & Metal Ball    & 17.85\%  & 100.00\% & N/A & N/A \\
Photo       & Mirror        & 96.42\%  & 100.00\% & N/A & N/A \\

\midrule

Credit                  & Metal Ball    & N/A    & N/A & 75.00\% & 96.42\% \\
Card        & Mirror        & N/A    & N/A & 71.42\% & 96.42\% \\

\midrule

\emph{Average}                 & Metal Ball    & 29.41\%      & 100.00\%  & 35.11\% & 91.67\% \\
\emph{Average} & Mirror        & 97.06\%      & 100.00\%  & 79.76\% & 95.83\% \\

\bottomrule

\end{tabular}
}
\label{tab:attack_results}
\vspace{-5mm}
\end{table}

Figure~\ref{fig:vis_attack} visualizes different types of sensitive information we can extract from reflective virtual objects.
We note that visually, a human user can easily identify sensitive information by inspecting the images displayed in row two. 
Table~\ref{tab:attack_results} shows that by leveraging automatic face \rec, our attack can recognize, on average, 29.41\% and 97.06\% faces automatically on the metal ball and mirror objects, respectively.
For automatic text \rec, we see an average success rate of 35.11\% and 79.76\% on the metal ball and mirror objects.
Further, via manual inspection of the renderings, we achieve 100\% face \rec rate on both metal ball and mirror objects, and up to 95.8\% \rec rate of textual information on the mirror object.

We make three key observations.
First, we can extract almost all sensitive information through manual inspection, despite the severe information distortions found on virtual object renderings.
Automatic recognition is very effective in extracting information from a flat object, suggesting the risk of large-scale automated reflection-based privacy attacks. 
Second, the manual inspection leads to a much higher information extraction success rate than the automatic recognition from the metal ball object.
We suspect that image distortion plays a key role in determining the success rate. 
Third, as image distortion can be caused by many factors such as the geometry of the scene, the geometry of the virtual object, and the viewing perspective, it can be challenging to devise an automatic unwrapping method.
However, to further improve the success rate of the automatic recognition, we believe one can resort to more accurate \rec methods as they are developed.  

\subsection{Effectiveness of the Defense}
\label{sub:defense_effectiveness}

We evaluate the visual quality impact and the visual perturbation effectiveness of our \defense on the reflective rendering.
For the visual quality impact, we compare the rendered objects to the ones generated by \emph{ARKit}, \emph{FusedAR}, and our \renderingdefense;
we also quantity the visual impacts using two image-based metrics (PSNR and SSIM) by calculating against the reflection renderings generated by \emph{FusedAR}. 
In summary, our results show that \defense has low visual impact when compared to the undefended privacy-risking renderings (up to 36db PSNR and 0.98 SSIM).
Further our \defense is effective in preserving reflection-based privacy, successfully decreasing the automatic information extraction rate to at most 8.8\%/5.4\% compared to 97\%/80\% when undefended.

\begin{figure}[t]
    \centering
\includegraphics[width=\linewidth]{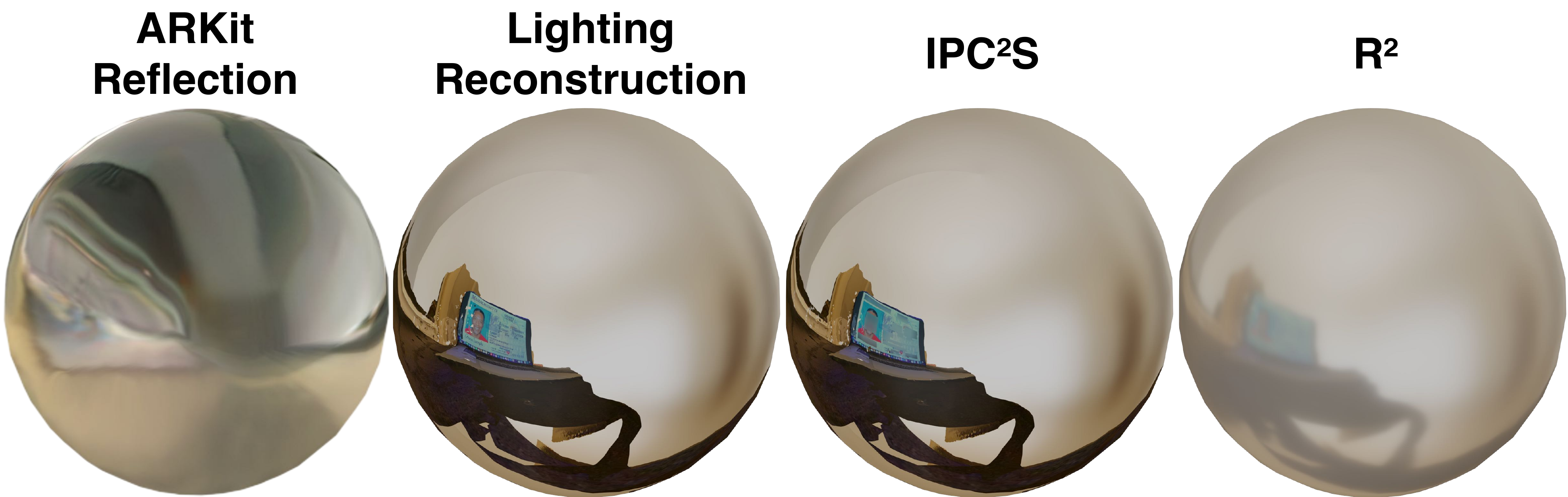} \vspace{-5mm}
    \caption{
        The low visual impact of our defense on reflection-based privacy.
        \textnormal{
        \defense achieves similar reflective visual quality compared to a re-implemented pipeline~\cite{zhao2022FusedAR}. 
        }
    }
    \label{fig:vis_defense_visuals}
\end{figure}

\begin{figure}[t]
    \centering

    \begin{subfigure}[b]{0.45\linewidth}
        \centering
\includegraphics[width=\linewidth]{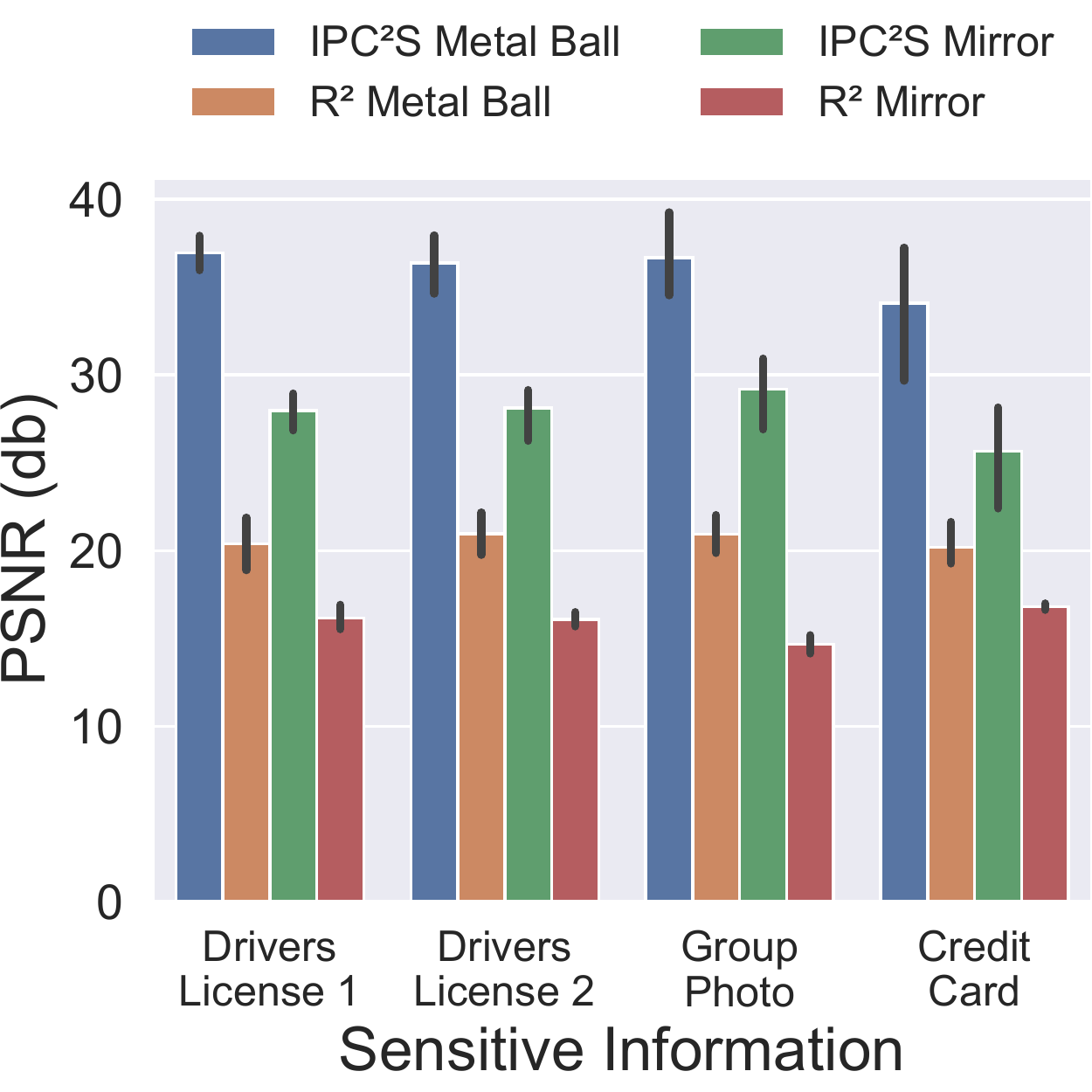}
        \vspace{-5mm}
        \caption{Rendering PSNR.
}
        \label{subfig:psnr}
    \end{subfigure}\quad
    \begin{subfigure}[b]{0.45\linewidth}
        \centering
\includegraphics[width=\linewidth]{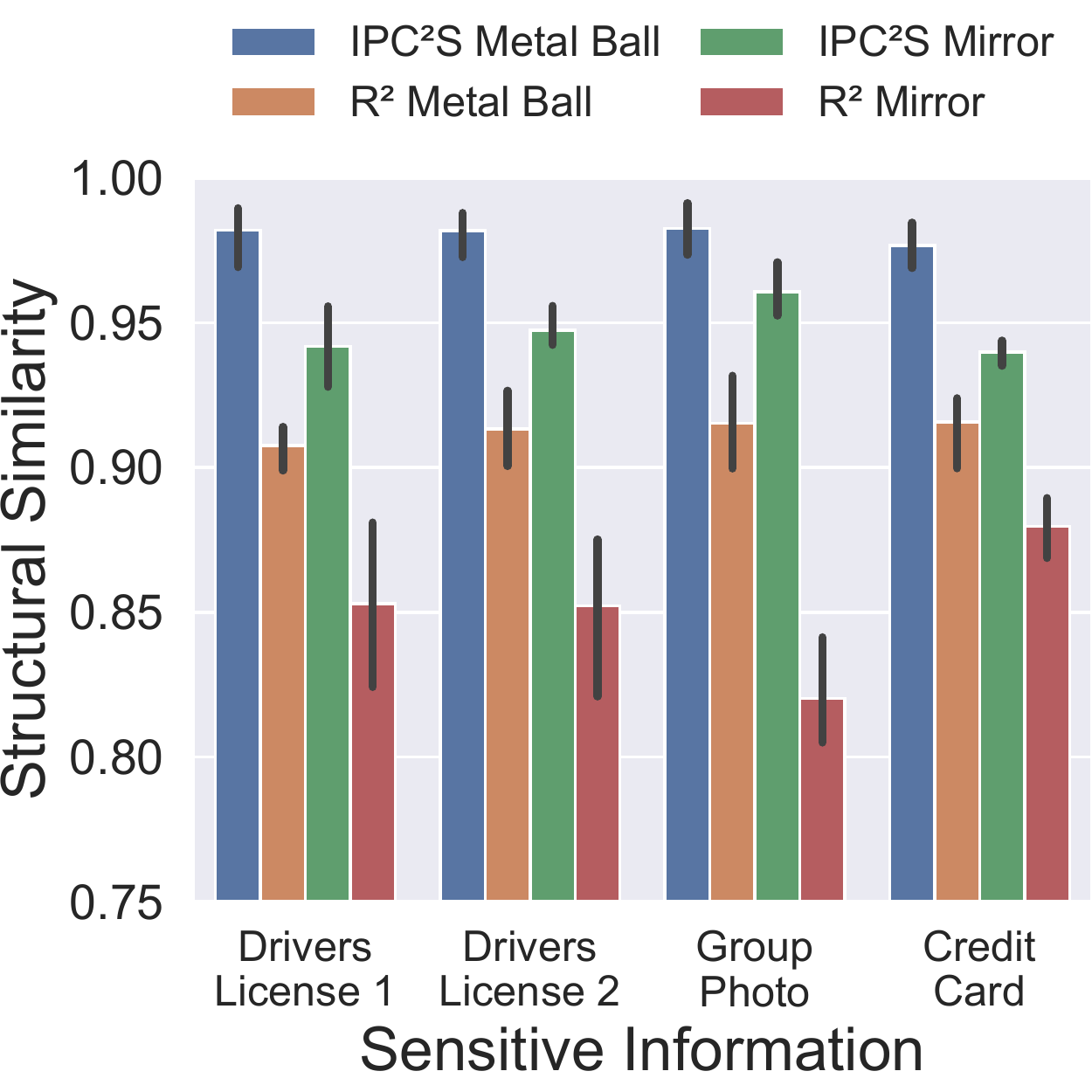}
        \vspace{-5mm}
        \caption{Rendering SSIM
}
        \label{subfig:ssim}
    \end{subfigure}

     \vspace{-4mm}
    
    \caption{
        Quantitative comparisons of our \defense and \renderingdefense.
        \textnormal{
        \defense achieves high PSNR and SSIM scores while \renderingdefense has at least 0.91 and 0.82 SSIM values. 
        }
    }
    \label{fig:defense_visual_impacts}
    \vspace{-3mm}
\end{figure}

\subsubsection{Visual Quality Impacts}
\label{subsub:defense_visual}
Figure~\ref{fig:vis_defense_visuals} visualizes the two reflective objects rendered with environment lighting information consisting of the driver's license 1 \footnote{The visualization of other rendering scenarios is omitted due to space limitations.}.
We first observe that our \defense can keep the detail of scene objects and geometries while only obfuscating the desirable information fields, compared to the rendering effects achieved with the re-implemented lighting reconstruction pipeline based on FusedAR (column two). 
The \renderingdefense also successfully obfuscates the sensitive information fields but lowers the overall visual quality. 
Specifically, with the reduction of the metallic property and introduction of roughness, reflective objects appear to have a matte-like looking.
As a reference, we also include the rendering generated using ARKit; we note that ARKit currently does not sufficiently support reflective renderings.
However, as newer lighting reconstruction techniques are adopted~\cite{zhao2022FusedAR,Somanath2020-of,prakash2019gleam} in commercial AR frameworks, the reflection-based privacy issue will become more prevalent.

Figure~\ref{fig:defense_visual_impacts} compares the rendering quality achieved by our \defense and \renderingdefense, against the undefended rendering. 
We first note that our \defense achieves up to on average 36.03db and 27.03db PSNR values for the metal ball and mirror objects, across all four scenes. Both the \defense and \renderingdefense achieve high SSIM scores for all tested rendering scenarios. 
Furthermore, we see that our \defense outperforms the \renderingdefense on both metal ball and mirror objects.
In particular, our \defense achieves 74.75\% and 7.43\% higher PSNR and SSIM than \renderingdefense on the metal ball object, as well as 74.86\% and 11.39\% higher PSNR and SSIM than \renderingdefense on the mirror object.
These results suggest that one should prioritize the use of \defense over \renderingdefense as much as possible to minimize the impact on visual quality.
As we described in \S~\ref{sec:defense}, we only fall back to \renderingdefense when the automated recognition accuracy and confidence fall below a certain threshold. 
As part of future work, we will investigate runtime policies to regulate the use of these two complementary defenses.

\setlength{\tabcolsep}{3pt}
\begin{table}[t]
\centering
\caption{
The low success rate when using our \defense.
\textnormal{
On average, this defense effectively decreases the success rate of automatic extraction to at most 8.8\%/5.4\% and manual extraction to at most 11.8\%/23.8\% for face/text information. 
}
}
\vspace{-3mm}
\small{ \begin{tabular}{lrrrrr}
\toprule
\textbf{Sensitive} & \textbf{Virtual} & \multicolumn{2}{r}{\textbf{Face Recognition}} & \multicolumn{2}{r}{\textbf{Text Recognition}} \\
\textbf{Info} & \textbf{Object} & \textbf{\textit{Automatic}} & \textbf{\textit{Manual}} & \textbf{\textit{Automatic}} & \textbf{\textit{Manual}}\\

\midrule

Driver's                    & Metal Ball    & 0.00\% & 0.00\% & 0.00\% & 16.67\% \\
License 1                   & Mirror        & 0.00\% & 0.00\% & 6.94\% & 18.06\% \\

\midrule

Driver's                    & Metal Ball    & 0.00\% & 0.00\% & 0.00\% & 30.88\% \\
License 2   & Mirror        & 0.00\% & 0.00\% & 5.88\% & 36.76\% \\

\midrule

Group                       & Metal Ball    & 0.00\%     & 14.29\% & N/A & N/A \\
Photo       & Mirror        & 10.71\%    & 14.29\% & N/A & N/A \\

\midrule

Credit                      & Metal Ball    & N/A    & N/A & 0.00\% & 3.57\% \\
Card        & Mirror        & N/A    & N/A & 0.00\% & 7.14\% \\

\midrule

\emph{Average}                     & Metal Ball    & 0.00\%     & 11.76\%  & 0.00\% & 20.24\% \\
\emph{Average}     & Mirror        & 8.82\%     & 11.76\%  & 5.36\% & 23.81\% \\

\bottomrule

\end{tabular}
}
\label{tab:defense_results}
\vspace{-2mm}
\end{table}

\subsubsection{Defense Success Rate}
\label{subsub:defense_success_rate}
Finally, we evaluate the effectiveness of our defenses following a similar methodology and metric as described in \S~\ref{subsec:attack_effectiveness}. 
Table~\ref{tab:defense_results} shows the information extraction success rate when using our \defense \footnote{\renderingdefense results are omitted as none of the sensitive information can be extracted.}.
First, we see that \defense can prevent at least 88.14\% and 76.19\% of face and text information extraction under manual inspection. 
This is in stark contrast to the 100\% human face and at least 92\% textual information extraction, if left undefended, as shown in Table~\ref{tab:attack_results}. 
Second, we show that our \defense is effective across all 32 rendering scenarios. 
For the automatic extraction, we can prevent all sensitive information from being leaked for the metal ball object and at least 91.2\% of information for the mirror object. 
Third, the larger difference in automatic extraction rates across the attack and the \defense subtly suggests that our point cloud-based, rather than image-based, defense design is more \rec model friendly.

\section{Related Work}

\para{AR/VR Security/Privacy}. With the growing popularity of AR/VR applications, the potential security and privacy issues caused by hybrid physical and virtual environments have recently emerged as a new research domain. 
Many research efforts have focused on the security/privacy implications of on-device sensors, as these sensors are increasingly utilized to capture sensitive user data or behaviors in order to build the immersive virtual environment~\cite{Sum-Security13,Fra-ACM14,Shi-mobicom21,Luo-NDSS20,Luo-VR22}.
Also, the sensitive nature of virtual objects constructed and presented in the AR/VR scenes has led to research efforts on deceptive virtual objects that mislead the users (i.e., the \emph{integrity} issue)~\cite{Kir-SP18,Kir-SP17}, as well as sensitive virtual objects~\cite{Tang-MM20} that can be abused by adversaries (i.e., the \emph{confidentiality} issue). 
Our work falls into the broad AR/VR privacy research: it differs from state-of-the-art works by targeting the non-conventional, reflective virtual objects rendered with views outside of any capturing devices (e.g., camera and sensors).

\para{Visual Privacy Protection.} Visual privacy protection has been a well-studied research topic for traditional multimedia, such as 2D images and videos~\cite{Rav-arxiv21}. The existing approaches to eliminating visual privacy leakage can be divided into two major categories. The first category aims to intervene/interfere with the sensors or scenes \emph{in the physical world} to prevent privacy-sensitive content from being captured in the first place~\cite{Zhu-mobicom17,Patel-ppvs09,Hinojosa_undated-qi}. The second category focuses on removing, replacing, or blurring sensitive objects \emph{in the virtual world} using computer vision techniques, such as image inpainting~\cite{Ber-TIP03,Cho-ICPR08}, body/face de-identification~\cite{Cor-cvpr21,Brk-cvprw17,Gaf-iccv19}, and image obfuscation~\cite{Fan-ICME19,Wang-arxiv20,Ren-ECCV18,Sun-CVPR18,Zhu-AIES20,Wu-mobicom21,Ye-MMSys22}. 
It is possible to apply existing techniques that remove reflection from an image~\cite{Liu-CVPR-2020,Liu_semantic_removeal,zhang2018single} by treating the rendered virtual object the same as its physical object counterpart. 
Though it is unclear how well existing techniques will work for more geometrically complex virtual objects or with distortion. 
In contrast, our defense mechanisms were designed with the knowledge of the inner working of the lighting reconstruction pipeline and thus can work in a more synergistic way.
Our work is inspired by the existing obfuscation-based approaches for visual privacy protection; 
in contrast to prior work, we target the privacy issues arising with the nascent development of photorealistic rendering~\cite{prakash2019gleam,pointar_eccv2020} in AR---a new multimedia that comes with brand new challenges including intricate visual quality, privacy, and performance trade-offs.

\vspace{-3mm}
\section{Conclusion and Future Work}

In this paper, we argued that unintentional privacy leakage can happen as augmented reality applications become popular. 
Specifically, sensitive information (such as human faces) can be leaked via reflective rendering---an integral part of photorealistic AR.
To underpin the importance of the reflection-based privacy issues, we showcased a simple attack leveraging a recently proposed multi-view lighting reconstruction~\cite{zhao2022FusedAR}. 
Our attack can successfully extract sensitive information under various rendering scenarios.
We also noted that such attacks are not specific to a particular lighting reconstruction method, and can also happen with existing commercial AR frameworks~\cite{arkit} and other academic works~\cite{prakash2019gleam}. 
The fundamental issue about this unintentional privacy leakage---in our example, between an AR content creator and a viewer---comes down to the seemingly conflicting goals of visual coherency and privacy. 

As explained previously, achieving visual coherency for AR objects requires an accurate understanding of the physical environment.
Based on current common practices to achieve visual coherency, it is inevitable that sensitive information will be captured and included as part of the environment scans. 
However, we showed that by carefully designing the defense pipeline in tandem with the lighting reconstruction, we can still achieve good visual coherency while preserving privacy. 
Specifically, we proposed two complementary defenses (\defense and \renderingdefense) to automatically obfuscate the sensitive information, even under dynamic environments.
As such, even if the sensitive information was captured as part of the AR sessions, it will not be subject to unintentional information leakage. 

Our proposed defense is far from complete---there are many unsolved challenges we plan to address.
For example, many objects can be considered private, and we only showcased the defense mechanism when considering human face and text information are sensitive. 
Currently, our work does not consider physical object reflection.
As AR deals with both virtual and physical spaces, we agree that an effective and complete system should consider both types of reflections to preserve user privacy.
Further, many practical scenarios exacerbate the problem, e.g., moving objects and varying environmental lighting. Robustly identifying private information with minimal human user involvement can be an exciting direction. Another direction is to devise better privacy-preserving techniques. Currently, we use a simple blurring technique to obfuscate sensitive information. We envision more sophisticated techniques such as automatically generating suitable replacements in real-time can achieve better visual quality.
We hope our study can improve the community's awareness of the need to support privacy-preserving reflection rendering.

\begin{acks}
We thank the anonymous reviewers for their constructive reviews. 
This work was supported in part by NSF Grants \#1815619, \#1912593, and \#2105564, and VMWare.
\end{acks}

\balance 
\bibliographystyle{ACM-Reference-Format}
\bibliography{AR-Sec-Ref,zotero,main}

\end{document}